\documentclass{PoS}

\newcommand{\be}{\begin{equation}}
\newcommand{\ee}{\end{equation}}

\title{Cosmic Neutrinos}

\ShortTitle{Cosmic Neutrinos}

\author{\speaker{Ofelia Pisanti}\\
        Dipartimento di Fisica E. Pancini, Universit\`a di Napoli Federico II, and INFN, Sezione di Napoli, Via Cintia, I-80126 Napoli, Italy.\\
        E-mail: \email{pisanti@na.infn.it}}

\abstract{Neutrinos are key astronomical messengers, because they are undeflected by magnetic field and unattenuated by electromagnetic interaction. After the first detection of extraterrestrial neutrinos in the TeV-PeV region by Neutrino Telescopes we are entering a new epoch where neutrino astronomy becomes possible. In this paper I briefly review the main issues concerning cosmological neutrinos and their experimental observation.}

\FullConference{The 19th International Workshop on Neutrinos from Accelerators-NUFACT2017\\
		25-30 September, 2017\\
		Uppsala University, Uppsala, Sweden}

\begin{document}

\section{The strong case for $\nu$ astronomy}

Direct exploration of the extragalactic high-energy universe above a few tens of TeV has been challenging because Cosmic Rays (CRs) are deflected by magnetic fields and confined to a $\sim 50$ Mpc horizon, while PeV gamma-rays are highly attenuated by diffuse light sources in the Universe. Neutrinos are, however, not deflected by galactic or extra-galactic magnetic field and do not interact with photons. Actually, neutrinos were immediately recognized as ideal cosmic messengers by one of their discoverer \cite{reines&cowan}, Frederik Reines, but also in the 60's by other theorists \cite{theor60_1,theor60_2}.

We expect high energy and ultra-high energy neutrinos produced by CRs interaction with cosmic matter, in or near their sources; cosmogenic or Greisen-Zatsepin-Kuzmin \cite{GZK_1,GZK_2} (GZK) neutrinos should also arise, in the region above $\sim$ PeV, from the interaction of CRs with Cosmic Microwave Background (CMB) and Extra-galactic Background Light (EBL).

In this brief review, not intending to be exhaustive, I will concentrate on some hot topics in this field.

\section{Neutrino signal in Neutrino Telescopes}

\begin{figure}[b]
\includegraphics[width=.6\textwidth]{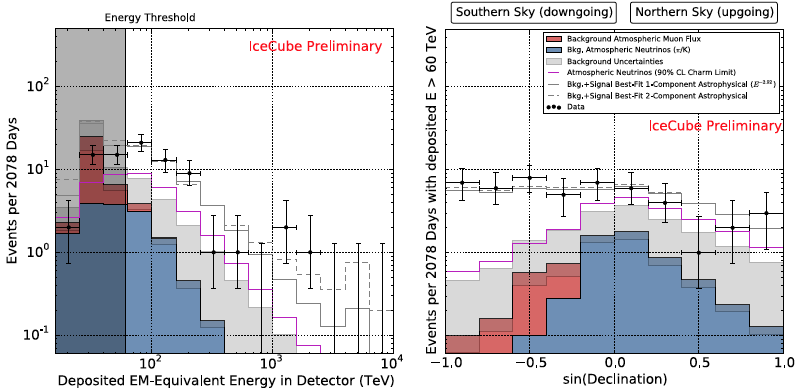}
\includegraphics[width=.4\textwidth]{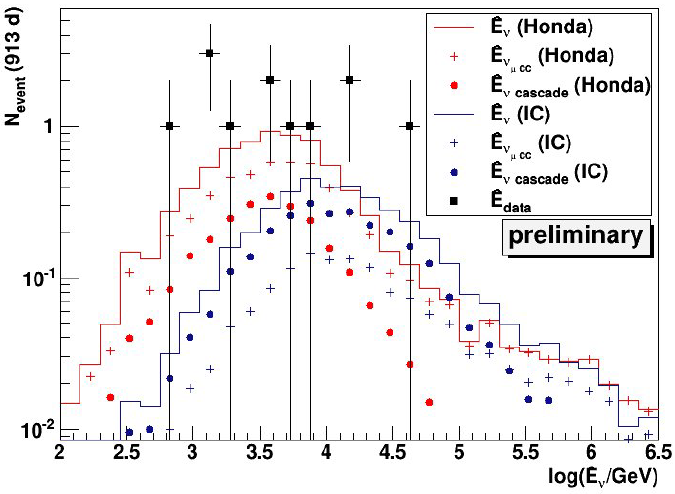}
\caption{Deposited energies and arrival directions of the observed IceCube events \cite{IceCubespectrum} (left) and energy distribution for the ANTARES events \cite{ANTARESspectrum} (right).}
\label{nusignal}
\end{figure}
In 2013 IceCube reported \cite{ICECUBEnu} the first detection of extraterrestrial neutrinos with energies between 10 TeV and 2 PeV, consistent with an extragalactic origin and fluxes of the order of the Waxman-Bahcall (WB) limit \cite{WB_1}. The hypothesis of a purely atmospheric explanation of the neutrino events can be rejected with 6.7$\sigma$ significance \cite{ICECUBEflux}. This signal was also confirmed by the experiment ANTARES, with a significance of 1.6$\sigma$ against the null cosmic flux assumption \cite{ANTARESflux}.

Neutrino data sample in IceCube or ANTARES is obtained with methods that aim to reject the atmospheric muon background. For example, IceCube samples upward-going and horizontal events ({\it upgoing} sample) or uses the outer part of the IceCube detector as an atmospheric muon veto detector ({\it starting} sample). On the other end, in ANTARES the distinction between atmospheric muons and neutrinos is accomplished by combining event angular estimates with the quality of event reconstruction by using energy-related variables. Usually, tracks have better than 1$^\circ$ resolution, cascades $\sim 15^\circ$ resolution. Then, distinction between atmospheric and cosmic $\nu$ events is achieved through energy. In fact, the spectrum is a combination of atmospheric (softer), prompt (less softer) and astrophysical (harder) neutrinos. The contribution of each of these components is estimated by multi-dimensional fits. Figure \ref{nusignal} shows the deposited energies and arrival directions of the observed IceCube events \cite{IceCubespectrum} and the energy distribution for the ANTARES events \cite{ANTARESspectrum}.
 
Observation of astrophysical $\nu$ by NT is a confirmation of the key role neutrinos could have, as probes of the non-thermal universe.

Which is the origin of these neutrinos? In the bottom-up scenario, neutrinos originate inside sources as the products of the interactions of charged CRs (typically protons) with photons or nucleons. Moreover, cosmogenic $\nu$'s are produced by the interaction of CRs with extra-galactic and galactic matter along their path to Earth. By assuming a given evolution with red-shift of the sources, one can have indications on the CRs production rate and then on secondary neutrinos and photons. So, it immediately appears that cosmic neutrino physics inherently has a multi-messenger character.

\section{Neutrinos in bottom-up or top-down scenarios}

\begin{figure}[t]
\begin{center}
\includegraphics[width=.42\textwidth]{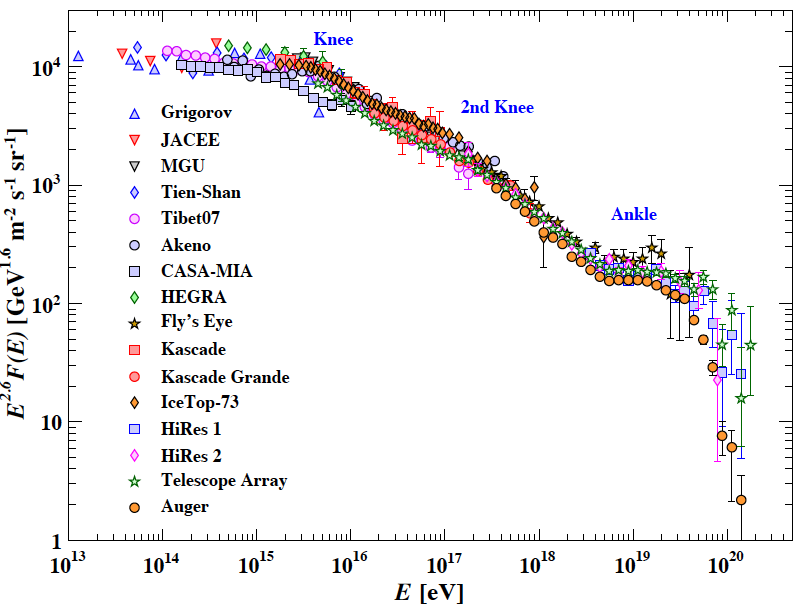}
\caption{All-particle CR spectrum as a function of the energy-per-nucleus from air-shower measurements \cite{PDG}.}
\label{CRspectrum}
\end{center}
\end{figure}
In the bottom-up scenario neutrinos come from primary CSs. As well known, CRs spectrum spans over more than 10 orders of magnitude (see figure \ref{CRspectrum}). CRs below the knee are likely produced by SuperNovae, but those at higher energies have a less certain origin. On the other side, Ultra-High Energy CRs (UHECRs) come almost certainly from extragalactic sources, because they cannot be confined by galactic magnetic fields. The production/acceleration of CRs could be achieved in a plethora of sources, like Active Galactic Nuclei (AGN), Gamma Ray Bursts, magnetars (neutron stars with petagauss surface magnetic fields), accretion shocks around cluster of galaxies, etc.. While pretty well known at low energy, mass composition of UHECRs is still debated. Two main scenarios can be considered: 1) pure proton composition (as suggested by the data analysis of the Telescope Array (TA) experiment \cite{TAmass}), the so called {\it dip} model \cite{dip_1,dip_2,dip_3}, where the ankle observed in the CR spectrum is an effect of the pair production by protons on the CMB, 2) mixed composition (as indicated by the results of the Paul Auger Observatory (PAO) \cite{PAOmass}), with protons at energies below $10^{18}$ eV and nuclei at higher energies.

Let us first consider production of neutrinos inside sources of production/acceleration of CRs. Connection of $\nu$ with CRs production rates is possible only for transparent sources. Photo-meson interaction of nucleons on photons produces pions (neutral and charged); then $\pi^\pm$'s decay in neutrinos:
\be
\pi^\pm \rightarrow e\nu_e\nu_\mu\nu_\mu ~~~~~~(\pi^\pm \rightarrow \mu\nu_\mu, ~\mu\rightarrow e\nu_e\nu_\mu)
\ee
Since pions carry on $\sim 20$\% of the nucleon energy and we have four particles in the final state, each neutrino carries on $\sim 5$\% of the primary CR energy. By considering the CR injection rate, $E^2 dN_p/(dEdt) \sim 10^{44}$ erg/(Mpc$^3$ yr), and a Fermi acceleration spectrum at the sources, $dN_p/dE \sim E^{-2}$, we obtain upper bounds on diffuse neutrino flux \cite{WB_1,WB_2,MPR} (magnetic field does not change so much the picture).

\begin{figure}[t]
\begin{center}
\includegraphics[width=.45\textwidth]{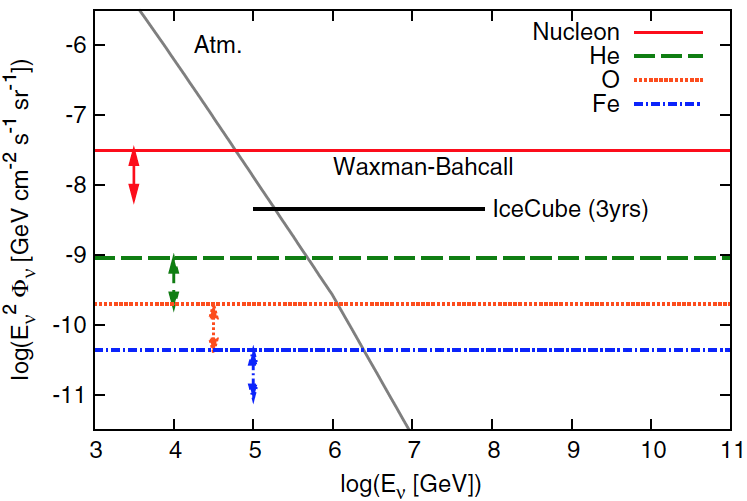}
\includegraphics[width=.45\textwidth]{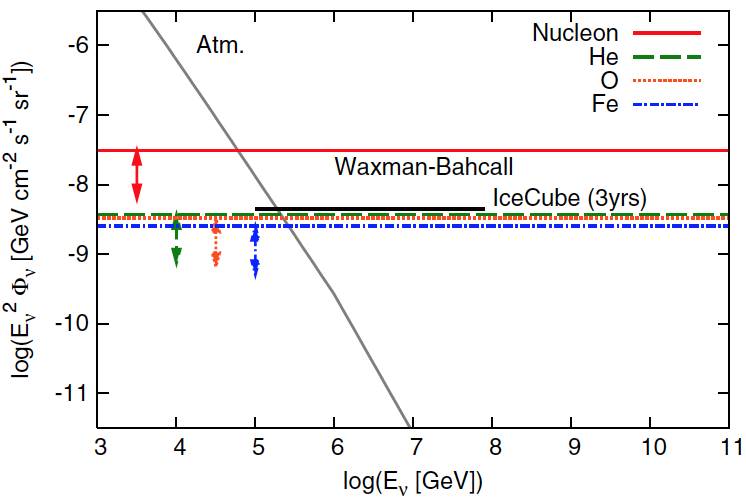}
\caption{Upper bounds for the $\nu$ background from UHECR photo-disintegration interactions in sources of different nuclei \cite{landmarks} compared with WB landmarks. Left: optical depth is used, right: effective optical depth is used. The arrows indicate the change for no redshift evolution.}
\label{murase&beacom}
\end{center}
\end{figure}
However, some authors claim that these upper bounds  should be considered as nominal scales ({\it landmarks}) and not limits \cite{landmarks}. Moreover, the same authors claim that they should be even lowered. In fact, for nuclei in radiation fields, the photo-disintegration process is even more important than the photo-meson process. Then, a heavy composition of UHECRs (according to the PAO results) would imply that more nuclei survive photo-disintegration. This can happen if less target photons are present in the sources, and then less neutrinos would be produced. Figure \ref{murase&beacom} shows the upper bounds for the $\nu$ background obtained for different nuclei, assuming the same injection spectrum as in \cite{WB_1,WB_2}, under the hypothesis that the photo-disintegration process happens mainly via the Giant Dipole Resonance. Left plot is obtained requiring optical depth less than 1, while right plot considers an effective optical depth. In both cases, small $\nu$ fluxes are obtained, at least one order of magnitude below the WB flux for UHE protons.

\begin{figure}[t]
\begin{center}
\includegraphics[width=.4\textwidth]{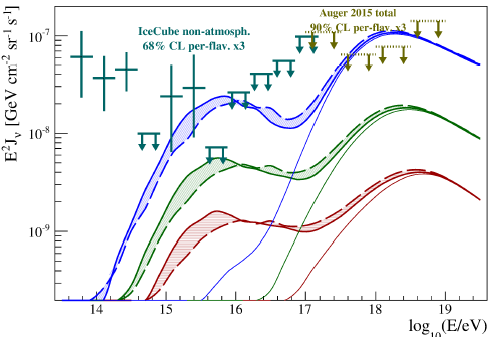}
\includegraphics[width=.401\textwidth]{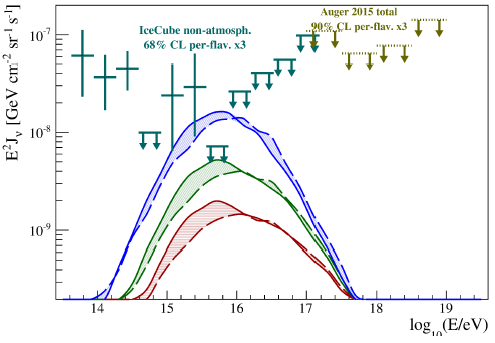}
\caption{Fluxes of neutrinos expected at Earth for different cosmological evolution of sources (from bottom to top: no evolution, star forming rate, and AGN-like evolution). Left: pure proton composition, right: mixed composition \cite{aloisio}.}
\label{aloisio}
\end{center}
\end{figure}

Even if the production of neutrinos near CR sources did not take place, cosmogenic neutrinos from CR interaction with CMB and EBL would be guaranteed. This is true under the assumption that CRs are extragalactic at the highest energies, which has been verified by the detection of a cutoff in the UHECR spectrum. In fact, due to the following processes:
\begin{itemize}
\item
pair photo-production: $p+\gamma\rightarrow p+e^++e^-$
\item
pion photo-production: $p+\gamma\rightarrow p+\pi^0$ or $p+\gamma\rightarrow n+\pi^+$
\item
nucleus photodisintegration: $(A,Z)+\gamma\rightarrow (A',Z')+N$
\end{itemize}
no protons with energies above $\sim 10^{18}$ eV can originate from $z>1$ ($\sim 50$ Mpc), the so called GZK feature \cite{GZK_1, GZK_2}. Since neutral pions decay in photons and charged pions and muons decay in neutrinos, which isotropically arrive to Earth, a multi-messenger approach can constrain UHECR sources and their distribution with red-shift. Propagation codes are typically used to study CR interaction with cosmic matter in such a way to compare the predicted flux of cosmogenic neutrinos with experimental limits. For example, in ref. \cite{aloisio} the authors use SimProp v2r2 \cite{SimProp} with injected spectra corresponding to the two composition scenarios considered before (pure proton or mixed), getting the results shown in figure \ref{aloisio}: while AGN-like evolution of sources results in a total neutrino flux in excess of the PAO and IceCube limits (left plot), extremely low neutrino fluxes at high energy are expected for the mixed composition scenario (right plot), because neutrinos come from pion production of nucleons in nuclei (then, with less energy per nucleon).

Alternative to the bottom-up is the top-down scenario, where neutrinos originate from the decay/annihilation of particles beyond the standard model (without violating WB-like bounds). A lot of models can be considered, which cannot be here exhaustively covered. I will give only an example. An intriguing possibility has been proposed \cite{chianese} that the slight excess in the range 40-200 TeV observed by IceCube, and at a lesser extent also by ANTARES, comes from the decay of dark matter with a mass of $\sim$ 400 TeV. Considering hadronic ($t+{\bar t}$) and leptonic ($\tau^++\tau^-$) channels, the best-fit values for this dark matter particle are $m_{DM}\simeq 500$ TeV and $\tau_{DM}\simeq 2.77\cdot 10^{27}$ s and $m_{DM}\simeq 400$ TeV and $\tau_{DM}\simeq 1.65\cdot 10^{28}$ s, respectively.

\section{Ultra high-energy neutrinos}

PAO \cite{PAO} in the southern hemisphere and TA \cite{TA} in the northern one are the two largest experiments for the detection of extensive air showers induced by CRs. Both are hybrid experiments with a surface array of Cherenkov/scintillators detectors combined with fluorescence detectors and, in their original configuration, they are sensible to CRs above $\sim 10^{18}$ eV.

\begin{figure}[t]
\begin{center}
\includegraphics[width=.4\textwidth]{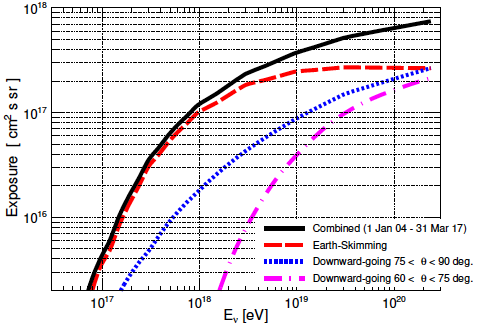}
\includegraphics[width=.4\textwidth]{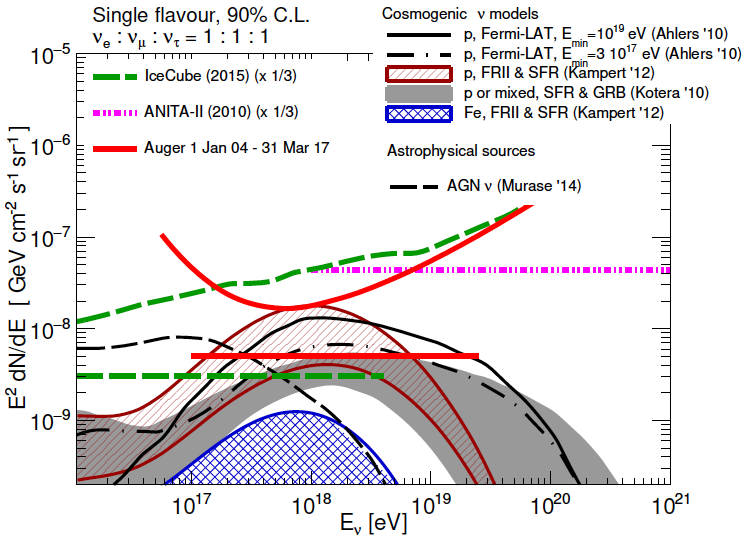}
\caption{Results of the PAO for cosmic neutrinos \cite{Zas}. Left: exposure as a function of neutrino energy, right: bounds (differential and integral) on the diffuse neutrino flux.}
\label{PAOnu}
\end{center}
\end{figure}

Neutrinos are disentangled from the bulk of ordinary CRs sampling inclined young showers, with a broad electromagnetic component. Bounds on the neutrino flux are obtained by a convolution of the exposure and the flux over neutrino energy. Figure \ref{PAOnu} (right) shows the integral bound (horizontal red line) obtained integrating the exposure with a conventional $E^{-2}$ spectrum, and the differential one (curved red line) calculated integrating the neutrino flux over consecutive energy bins. Note that, as evident from left plot in figure \ref{PAOnu}, earth-skimming neutrinos dominate the exposure in spite of the reduced solid angle to which the detector is sensitive. As shown by the plot, PAO bound already starts constraining some cosmogenic or astrophysical models.

The IceCube signal isotropy implies an extragalactic origin for detected neutrinos. Then, if IceCube neutrinos come from $pp$/$p\gamma$ interactions, it is possible that they point to the same sources of UHECRs. Now, both PAO and TA observe the GZK suppression, which would imply that the sources have distances shorter than $\sim$ 200 Mpc. However: 1) no statistically significant small-scale anisotropy and correlation with nearby astrophysical sources have been established until now; 2) IceCube neutrino energies are much smaller than those of UHECRs.

There are arguments that can explain the previous points: 1) we can imagine sources that produce PeV neutrinos and not UHECRs, 2) differently from UHECRs, neutrinos come from sources at any distance, 3) due to the time delay induced by ambient magnetic fields on charged CRs, neutrinos could not be detected in time coincidence with the other primaries, unless we consider continuous emitting sources.

\begin{figure}[b]
\begin{center}
\includegraphics[width=.85\textwidth]{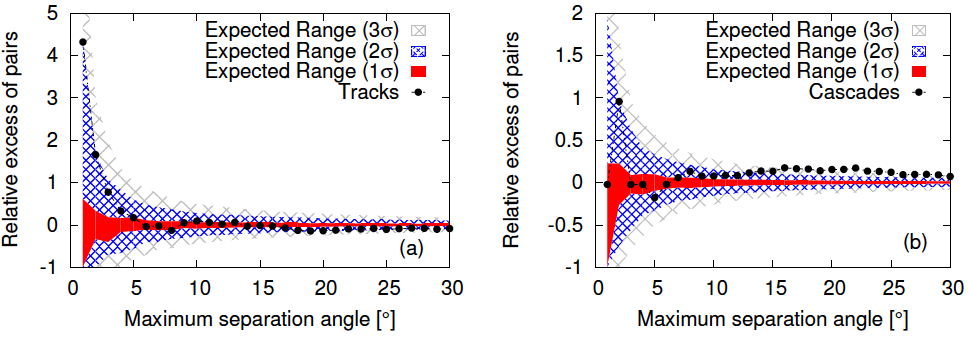}
\caption{Fractional excess relative to the expectation for an isotropic CR distribution as a function of the angular separation between the IceCube neutrino and cosmic-ray pairs \cite{correlation}. Left: track events, right: cascade events.}
\label{correlation}
\end{center}
\end{figure}
In order to clarify the issue, a combined analysis \cite{correlation} was made in search of correlation between the arrival directions of UHECRs detected by PAO and TA, and of 39 high-energy cascade events and 16 high-energy tracks (HESE events plus high energy through-going muons) observed by IceCube. The results are compatible at 2$\sigma$ with the ones from an isotropic CR distribution, even if small excesses of tracks at small angle separation ($1^\circ$) and of cascades at intermediate angle ($22^\circ$) are observed (see figure \ref{correlation}) (possibly due to the concentration of $\nu$ and CR events near the Super-Galactic plane and at the TA {\it hot spot}). Further studies are needed also in view of some open issue, like the energy calibration in the PAO-TA experiments and the uncertainties from magnetic deflection and mass composition of UHECRs.

\section{A look into the future}

One very interesting technique is the radio detection, that is detection of radio impulse Cheren\-kov emission due to the dipole of negative and positive charges in the geomagnetic field or the Askaryan effect. Several in progress or future experiments are using or plan to use this technique \cite{AERA,ANITA,ARA,ARIANNA,GNO}.

Besides new experimental techniques, several experiments plan to upgrade their configuration for focusing on promising issues and new experiments are in their R\&D phase. IceCube$-$Gen2 \cite{Gen2} aims in increasing the detector volume by a factor of 10. Auger-Prime \cite{AugerPrime} upgrade consists in new plastic scintillator detectors on top and an additional PMT installed in the water-Cherenkov detectors of the PAO surface array, and more powerful electronics. KM3NeT, the network of deep-sea NTs in the Mediterranean Sea, is building two experiments with different granularity, with the objectives of the discovery and subsequent observation of high-energy neutrino sources in the Universe (ARCA \cite{ARCA}) and the determination of the mass hierarchy of neutrinos (ORCA \cite{ORCA}). The Gigaton Volume Detector GVD in Lake Baikal \cite{Baikal-GVD} starts from an initial 0.4 km$^3$ effective volume and then will surpass the cubic kilometer benchmark. JEM$-$EUSO \cite{JEMEUSO} will install a wide field of view fluorescence telescope onboard of the International Space Station for detecting from space fluorescence and Cerenkov light from $\nu$ initiated showers.

\section{Conclusions}

A diffuse flux of extra-galactic neutrinos around PeV energies has been observed by NT experiments at the level of existing theoretical upper bounds, for which atmospheric neutrino explanation is excluded at more than 5$\sigma$. On the other side, we still wait for UHE neutrinos at giant air shower arrays. In any case, we are entering a new epoch, where neutrino astronomy replaces cosmic neutrino discovery. To be able to pursue this exciting goal, new experimental techniques or upgrade of existing experiments are necessary. However, the recent announcements of joint gravitational waves and high-energy electromagnetic observations suggests that neutrino astronomy has to be a part of a more complex scenario where different messengers are put together for enhancing the possibility of understanding the universe.


\begin{thebibliography}{99}

\bibitem{reines&cowan}
F. Reines, C. L. Cowan Jr, \emph{The neutrino}, \emph{Nature} {\bf 178} (1956) 446.

\bibitem{theor60_1}
K. Greisen, \emph{Cosmic ray showers}, \emph{Ann. Rev. Nucl. Part. Sci.} {\bf 10} (1960) 63.

\bibitem{theor60_2}
M.A.Markov in \emph{Proceedings of the 1960 International Conference on High Energy Physics}, E.C.G. Sudarshan, J.H.Tinlot and A.C.Melissinos Editors (1960) 578.

\bibitem{GZK_1}
K. Greisen, \emph{End to the cosmic ray spectrum?}, \emph{Phys. Rev. Lett.} {\bf 16} (1966) 748.

\bibitem{GZK_2}
G.T. Zatsepin and V.A. Kuzmin, \emph{Upper limit of the spectrum of cosmic rays}, \emph{JETP Lett.} {\bf 4} (1966) 78.

\bibitem{ICECUBEnu}
M. G. Aartsen et al. [IceCube Collab.], \emph{Evidence for High-Energy Extraterrestrial Neutrinos at the IceCube Detector}, \emph{Science} {\bf 342} (2013) 1242856.

\bibitem{WB_1}
J.N. Bahcall, E. Waxman, \emph{High-energy neutrinos from astrophysical sources: An Upper bound}, \emph{Phys. Rev. D} {\bf 59} (1999) 023002.

\bibitem{ICECUBEflux}
The IceCube Collab., \emph{A Measurement of the Diffuse Astrophysical Muon Neutrino Flux Using Eight Years of IceCube Data}, in proceedings of the \emph{35th International Cosmic Ray Conference (ICRC2017)}, \pos{PoS(ICRC2017)1005}.

\bibitem{ANTARESflux}
The ANTARES Collab., \emph{All-flavor search for a diffuse flux of cosmic neutrinos with 9 years of ANTARES data}, {\tt arXiv:1711.07212}.

\bibitem{IceCubespectrum}
C. Kopper, on behalf of the IceCube Collab., \emph{Observation of Astrophysical Neutrinos in Six Years of IceCube Data}, in proceedings of the \emph{35th International Cosmic Ray Conference (ICRC2017)}, \pos{PoS(ICRC2017)981}.

\bibitem{ANTARESspectrum}
S. Hallmann, on behalf of the ANTARES Collab., \emph{Search for a diffuse cosmic neutrino flux with ANTARES using track and cascade events}, in proceedings of the \emph{34th International Cosmic Ray Conference (ICRC2015)}, \pos{PoS(ICRC2015)1065}.

\bibitem{PDG}
C. Patrignani {\it et al.} (Particle Data Group), \emph{Cosmic rays} review, \emph{Chin. Phys. C}, {\bf 40} (2016) 100001 and 2017 update.

\bibitem{TAmass}
D. Ikeda, W. Hanlon, on behalf of the Telescope Array Collab., \emph{Hybrid Measurement of the Energy Spectrum and Composition of Ultra-High Energy Cosmic Rays by the Telescope Array}, in proceedings of the \emph{35th International Cosmic Ray Conference (ICRC2017)}, \pos{PoS(ICRC2017)515}.

\bibitem{dip_1}
V. S. Berezinsky and S. I. Grigorieva, \emph{A bump in the ultra-high energy cosmic ray spectrum}, \emph{Astron. Astroph.} {\bf 199} (1988) 1.

\bibitem{dip_2}
V. Berezinsky, A. Z. Gazizov and S. I. Grigorieva, \emph{On astrophysical solution to ultrahigh energy cosmic rays}, \emph{Phys. Rev. D} {\bf 74} (2006) 043005.

\bibitem{dip_3}
V. Berezinsky, A. Z. Gazizov and S. I. Grigorieva, \emph {Dip in UHECR spectrum as signature of proton interaction with CMB}, \emph{Phys. Lett. B} {\bf 612} (2005) 147.

\bibitem{PAOmass}
J. Bellido, on behalf of the Pierre Auger Collab., \emph{Depth of maximum of air-shower profiles at the Pierre Auger Observatory: Measurements above $10^{17.2}$ eV and Composition Implications}, in proceedings of the \emph{35th International Cosmic Ray Conference (ICRC2017)}, \pos{PoS(ICRC2017)506}.

\bibitem{WB_2}
J.N. Bahcall, E. Waxman, \emph{High-energy astrophysical neutrinos: The Upper bound is robust}, \emph{Phys. Rev. D} {\bf 64} (2001) 023002.

\bibitem{MPR}
K. Mannheim, R.J. Protheroe, J.P. Rachen, \emph{On the cosmic ray bound for models of extragalactic neutrino production}, \emph{Phys. Rev. D} {\bf 63} (2001) 023003.

\bibitem{landmarks}
K. Murase and J.F. Beacom, \emph{Neutrino background flux from sources of ultrahigh-energy cosmic-ray nuclei}, \emph{Phys. Rev. D} {\bf 81} (2010) 123001.

\bibitem{aloisio}
R. Aloisio, D. Boncioli, A. di Matteo, A.F. Grillo, S. Petrera, F. Salamida, \emph{Cosmogenic neutrinos and ultra-high energy cosmic ray models}, \emph{JCAP} {\bf 10} (2015) 006.

\bibitem{SimProp}
R. Aloisio, D. Boncioli, A.F. Grillo, S. Petrera and F. Salamida, \emph{SimProp: a simulation code for
ultra high energy cosmic ray propagation}, \emph{JCAP} {\bf 10} (2012) 007.

\bibitem{chianese}
M. Chianese, G. Miele, S. Morisi, \emph{Interpreting IceCube 6-year HESE data as an evidence for hundred TeV decaying Dark Matter}, \emph{Phys. Lett. B} {\bf 773} (2017) 591-595.

\bibitem{PAO}
J. Abraham, {\it et al.}, \emph{Properties and performance of the prototype instrument for the Pierre Auger Observatory}, \emph{Nucl. Instr. Meth. Phys. Res. A} {\bf 523} (2004) 50.

\bibitem{TA}
Y. Tsunesada, on behalf of the Telescope Array Collab., \emph{Highlights from Telescope Array}, in proceedings of the \emph{32th International Cosmic Ray Conference (ICRC2011)} (2011) 67 [{\tt arXiv:1111.2507}].

\bibitem{Zas}
E. Zas, \emph{Searches for neutrino fluxes in the EeV regime with the Pierre Auger Observatory}, in proceedings of the \emph{35th International Cosmic Ray Conference (ICRC2017)}, \pos{PoS(ICRC2017)972}.

\bibitem{correlation}
IceCube and Pierre Auger and Telescope Array Collab.s, \emph{Search for correlations between the arrival directions of IceCube neutrino events and ultrahigh-energy cosmic rays detected by the Pierre Auger Observatory and the Telescope Array}, \emph{JCAP} {\bf 01} (2016) 037.

\bibitem{AERA}
J. Schulz, on behalf of the Pierre Auger Collab., \emph{Status and Prospects of the Auger Engineering Radio Array}, in proceedings of the \emph{34th International Cosmic Ray Conference (ICRC2015)}, \pos{PoS(ICRC2015)615}.

\bibitem{ANITA}
P. W. Gorham {\it et al.}, \emph{Observational constraints on the ultrahigh energy cosmic neutrino flux from the second flight of the ANITA experiment}, \emph{Phys. Rev. D} {\bf 82} (2010) 022004.

\bibitem{ARA}
P. Allison {\it et al.}, \emph{Performance of two Askaryan Radio Array stations and first results in the search for ultrahigh energy neutrinos}, \emph{Phys. Rev. D} {\bf 93} (2016) 082003.

\bibitem{ARIANNA}
S.W. Barwick {\it et al.}, \emph{Radio detection of air showers with the ARIANNA experiment on the Ross Ice Shelf}, {\tt arXiv:1612.04473}.

\bibitem{GNO}
S. Wissel {\it et al.}, \emph{Site Characterization and Detector Development for the Greenland Neutrino Observatory}, in proceedings of the \emph{34th International Cosmic Ray Conference (ICRC2015)}, \pos{PoS(ICRC2015)1150}.

\bibitem{Gen2}
M.G. Aartsen {\it et al.}, in proceedings of the \emph{Frontier Research in Astrophysics - II (FRAPWS2016)} [{\tt arXiv:1412.5106}].

\bibitem{AugerPrime}
D. Martello, on behalf of the Pierre Auger Collab., \emph{The Pierre Auger Observatory Upgrade}, in proceedings of the \emph{35th International Cosmic Ray Conference (ICRC2017)}, \pos{PoS(ICRC2017)383}.

\bibitem{ARCA}
P. Piattelli, on behalf of the KM3NeT Collab., \emph{All-flavour high-energy neutrino astronomy with KM3NeT/ARCA}, in proceedings of the \emph{34th International Cosmic Ray Conference (ICRC2015)}, \pos{PoS(ICRC2015)1158}.

\bibitem{ORCA}
J. Brunner, on behalf of the KM3NeT Collab., \emph{KM3NeT - ORCA: Measuring neutrino oscillations and the mass hierarchy in the Mediterranean}, in proceedings of the \emph{34th International Cosmic Ray Conference (ICRC2015)}, \pos{PoS(ICRC2015)1140}.

\bibitem{Baikal-GVD}
A. Avrorin {\it et al.} [Baikal Collab.], \emph{The prototyping/early construction phase of the BAIKAL-GVD project}, \emph{Nucl. Instrum. Meth. A} {\bf 742} (2014) 82.

\bibitem{JEMEUSO}
Y. Takahashi {\it et al.}, \emph{The JEM-EUSO mission}, \emph{New J. Phys.} {\bf 11} (2009) 065009.

\end{thebibliography}
\end{document}